\documentclass[amsmath,amssymb,superscriptaddress,nobalancelastpage,prl,twocolumn,longbibliography]{revtex4-1}

\usepackage{graphicx}
\usepackage{varioref}
\usepackage{xr-hyper}
\usepackage{xcolor}
\usepackage{hyperref}
\usepackage{ulem}
\usepackage{braket}
\usepackage{soul}
\newcommand{\Ca}{Ca$_{2}$RuO$_4$}

\newcommand{\dxy}{$d_{xy}$}
\newcommand{\dxz}{$d_{xz}$}
\newcommand{\dyz}{$d_{yz}$}

\newcommand{\A}{A}
\newcommand{\B}{B}
\newcommand{\C}{C}
\newcommand{\D}{D}

\begin{document}

\title{Spin-Orbital Excitations in \Ca\ Revealed by Resonant Inelastic X-ray Scattering}
    
\author{L.~Das}
\affiliation{Physik-Institut, Universit\"{a}t Z\"{u}rich, Winterthurerstrasse 190, CH-8057 Z\"{u}rich, Switzerland}
  
\author{F.~Forte}
\affiliation{CNR-SPIN, I-84084 Fisciano, Salerno, Italy}
\affiliation{Dipartimento di Fisica ``E.R.~Caianiello", Universit\`{a} di Salerno, I-84084 Fisciano, Salerno, Italy}
  
\author{R.~Fittipaldi}
\affiliation{CNR-SPIN, I-84084 Fisciano, Salerno, Italy}
\affiliation{Dipartimento di Fisica ``E.R.~Caianiello", Universit\`{a} di Salerno, I-84084 Fisciano, Salerno, Italy}

\author{C.~G.~Fatuzzo}
\affiliation{Institute of Physics, \'{E}cole Polytechnique Fed\'{e}rale de Lausanne (EPFL), CH-1015 Lausanne, Switzerland}

\author{V.~Granata}
\affiliation{CNR-SPIN, I-84084 Fisciano, Salerno, Italy}
\affiliation{Dipartimento di Fisica ``E.R.~Caianiello", Universit\`{a} di Salerno, I-84084 Fisciano, Salerno, Italy}

\author{O.~Ivashko}
\affiliation{Physik-Institut, Universit\"{a}t Z\"{u}rich, Winterthurerstrasse 190, CH-8057 Z\"{u}rich, Switzerland}
    
    \author{M.~Horio}
\affiliation{Physik-Institut, Universit\"{a}t Z\"{u}rich, Winterthurerstrasse 190, CH-8057 Z\"{u}rich, Switzerland}
    
\author{F.~Schindler}
\affiliation{Physik-Institut, Universit\"{a}t Z\"{u}rich, Winterthurerstrasse 190, CH-8057 Z\"{u}rich, Switzerland}
       
\author{M.~Dantz}
\affiliation{Swiss Light Source, Paul Scherrer Institut, CH-5232 Villigen PSI, Switzerland}

\author{Yi Tseng} 
\affiliation{Swiss Light Source, Paul Scherrer Institut, CH-5232 Villigen PSI, Switzerland}

\author{D. McNally}
\affiliation{Swiss Light Source, Paul Scherrer Institut, CH-5232 Villigen PSI, Switzerland}

\author{H.~M.~R\o{}nnow}
\affiliation{Institute of Physics, \'{E}cole Polytechnique Fed\'{e}rale de Lausanne (EPFL), CH-1015 Lausanne, Switzerland}

\author{W. Wan}
\affiliation{Department of Physics, Technical University of Denmark, DK-2800 Kongens Lyngby, Denmark}

\author{N.~B. Christensen}
\affiliation{Department of Physics, Technical University of Denmark, DK-2800 Kongens Lyngby, Denmark}
  
\author {J.~Pelliciari}  
\altaffiliation{Present address: Department of Physics, Massachusetts Institute of Technology, Cambridge,MA 02139, USA}
\affiliation {Swiss Light Source, Paul Scherrer Institut, CH-5232 Villigen PSI, Switzerland}  
  
\author{P.~Olalde-Velasco}
\altaffiliation{Present address: Instituto de Fisica, Benemerita Universidad Autonoma de Puebla, Apdo. Postal J-48, Puebla, Puebla 72570, Mexico}
\affiliation{Swiss Light Source, Paul Scherrer Institut, CH-5232 Villigen PSI, Switzerland} 

\author{N.~Kikugawa}
\affiliation{National Institute for Materials Science, 1-2-1 Sengen,  Tsukuba,  305-0047  Japan} 
\affiliation{National High Magnetic Field Laboratory, Tallahassee, Florida 32310, USA}


\author{T.~Neupert}
\affiliation{Physik-Institut, Universit\"{a}t Z\"{u}rich, Winterthurerstrasse 190, CH-8057 Z\"{u}rich, Switzerland}

\author{A.~Vecchione}
\affiliation{CNR-SPIN, I-84084 Fisciano, Salerno, Italy}
\affiliation{Dipartimento di Fisica ``E.R.~Caianiello", Universit\`{a} di Salerno, I-84084 Fisciano, Salerno, Italy}
 
\author{T.~Schmitt}
\affiliation{Swiss Light Source, Paul Scherrer Institut, CH-5232 Villigen PSI, Switzerland}

\author{M.~Cuoco}
\affiliation{CNR-SPIN, I-84084 Fisciano, Salerno, Italy}
\affiliation{Dipartimento di Fisica ``E.R.~Caianiello", Universit\`{a} di Salerno, I-84084 Fisciano, Salerno, Italy}

\author{J.~Chang}
\affiliation{Physik-Institut, Universit\"{a}t Z\"{u}rich, Winterthurerstrasse 190, CH-8057 Z\"{u}rich, Switzerland}

\begin{abstract}
The strongly correlated insulator \Ca\ is considered as a paradigmatic realization of both spin-orbital physics and a band-Mott insulating phase, characterized by  orbitally selective coexistence of a band and a Mott gap.
We present a high-resolution oxygen $K$-edge resonant inelastic X-ray scattering study of the antiferromagnetic Mott insulating state of \Ca. A set of low-energy ($\sim$80 and 400~meV) and high-energy ($\sim1.3$ and 2.2~eV)  excitations 
are reported that show strong  incident light polarization dependence. 
Our results strongly support a spin-orbit coupled  band-Mott scenario and explore in detail the nature of its exotic excitations.
Guided by theoretical modelling, we interpret the low-energy excitations as a result of composite spin-orbital excitations. 
Their nature unveil the intricate interplay of crystal-field splitting and spin-orbit coupling in the band-Mott scenario. The high-energy excitations correspond to intra-atomic singlet-triplet transitions at an energy scale set by the Hund's coupling.
Our findings give a unifying picture of the spin and orbital excitations in the band-Mott insulator \Ca.

\end{abstract}

\maketitle
  
\textit{Introduction.}  
Spin-orbit coupling (SOC) is a central thread in the search for novel 
quantum material physics~\cite{krempaARCMP2014}. A particularly promising avenue is the combination 
of SOC and strong electron correlations in multi-orbital systems.
This scenario is realized in heavy transition metal oxides composed of $4d$ and $5d$ elements.
Iridium-oxides (iridates) such as Sr$_2$IrO$_4$ are prime examples of systems where 
SOC plays a defining role in shaping the Mott insulating ground state~\cite{bjkimPRL2008}.
In fact, spin-orbit entanglement essentially outplays the effectiveness of 
the usually influential crystal field $\delta$. 
Of equal interest is the complex regime where SOC and crystal field energy 
scales are comparable. Here \Ca\ is a topical material that displays a wealth of physical properties.
A record high non-superconducting diamagnetic response has, for example, been reported recently~\cite{Sow1084}. 
Superconductivity emerges in  strained films~\cite{Nobukane2017}  or upon application of hydrostatic pressure to bulk  crystals~\cite{AlirezaJPCM10}. 
Neutron and Raman scattering experiments 
have demonstrated both phase and amplitude spin-excitation modes 
consistent with the existence of a spin-orbit exciton~\cite{KunkemollerPRL2015,jain2017higgs,Souliou2017}.
Moreover, measurements of the paramagnetic insulating band structure~\cite{SutterNatComm2017a} were interpreted in favor of an orbitally differentiated band-Mott 
insulating ground state~\cite{LiebschPRL2007,GorelovPRL2010}.
This rich phenomenology of \Ca\ is a manifestation of the interplay between 
multiple energy scales, specifically, the Coulomb interaction $U$, the Hund's coupling $J_\mathrm{H}$, 
the crystal field splitting $\delta$ and SOC $\lambda$.
In particular, a tendency towards an orbital selective Mott state is expected to be driven by the Hund's coupling~\cite{Georges2013}. 
Furthermore, the band-Mott scenario is triggered by a crystal field that renders the 
\dxy\ orbital band insulating, such that the resulting half-filled \dxz,~\dyz\ band 
is undergoing a conventional Mott transition driven by the Coulomb interaction~\cite{GorelovPRL2010}.

The low-energy electronic excitations of \Ca\ have been interpreted within 
an exciton picture where SOC enters as an important parameter~\cite{Khaliullin2017}.
A similar framework has been applied to layered iridates -- with a 5/6 filled $t_{2g}$ shell -- where a
$J_{\mathrm{eff}}=1/2$ quasiparticle
emerges from strong SOC. The existence of this quasiparticle has been 
confirmed by detailed resonant inelastic X-ray scattering (RIXS) studies of both spin and orbital excitations~\cite{KimPRL2012,KimNatComm14}. 
For \Ca, with modest spin-orbit coupling strength, studies of the spin excitations 
have been interpreted as evidence for a similar composite $J_{\mathrm{eff}}=1$ quasiparticle~\cite{KhaliullinPRL2013,AkbariPRB2014}.
However, the full manifold of the low-lying spin-orbital excitations of \Ca\ has not yet been observed.
  The possibility to detect Ru $d$-orbital excitations through the oxygen $K$-edge~\cite{BisogniPRB12,LuPRB2018}, offers a unique opportunity
  in the case of ruthenates, where direct $L$-edge RIXS is not yet available for high-resolution measurements. Moreover, spin-orbital excitations 
  are mostly inaccessible to neutron scattering.

\begin{figure*}
	\centering
\includegraphics[width=1\textwidth]{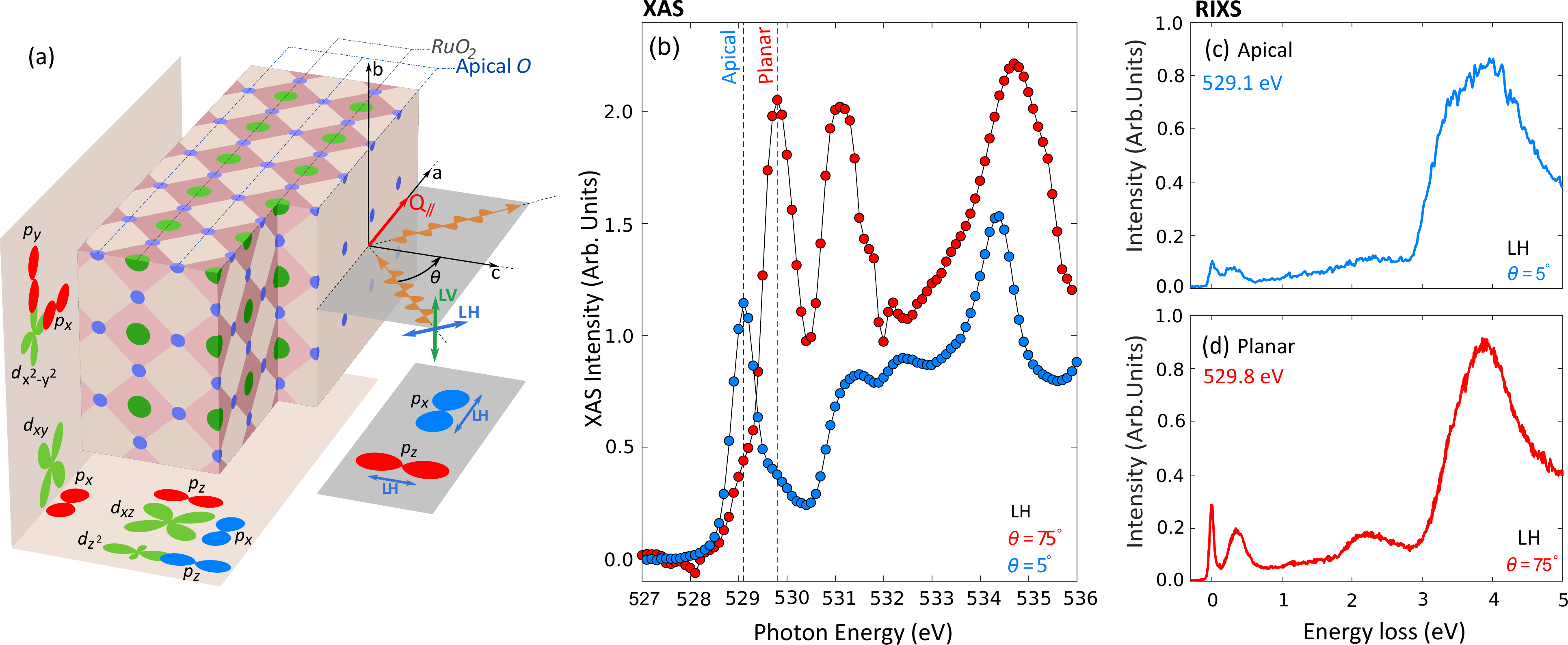}
\caption{(color online) (a) RIXS geometry with respect to the crystal lattice of \Ca\ is displayed schematically.  
Ruthenium and oxygen sites are shown with filled green and blue circles respectively. The variable incident angle $\theta$ is defined with respect to 
the RuO$_2$ and apical oxygen planes. Using LV and LH polarized light, for different $\theta$, sensitivity to either 
oxygen $p_x$, $p_y$ or $p_z$ orbitals can be obtained. These oxygen orbitals in turn hybridize with different unoccupied $t_{2g}$ and $e_g$ states on the ruthenium site. 
(b--d) XAS and RIXS spectra recorded with linear horizontal light for near grazing and normal incident light conditions as indicated. 
(b) Background subtracted  X-ray absorption spectra recorded with settings that either optimize the apical or planar oxygen K-edge resonances as indicated by the dashed vertical lines. (c) and (d) display RIXS spectra measured at the planar and apical  oxygen $K$-edges.} 
\label{fig:fig1n}
 \end{figure*}
 
Here we present an oxygen $K$-edge RIXS study of \Ca\ focusing on the magnetically ordered phase. 
Two low energy excitations (80 and 400 meV) and two high energy excitations (1.3 and 2.2 eV) are identified. Light-polarization analysis yields insight to the internal orbital character of these excitations. A detailed analysis of the 400 meV 
excitation uncovered a weak dispersion, consistent with a propagating nature.
 
In contrast, the high-energy excitations, are closely linked to the Hund's coupling energy scale $J_\mathrm{H}$. 
The excitations reported here on \Ca\ are  unique 
features of 
(1) a band-Mott insulating phase controlled by Hund's coupling and Coulomb interactions, and
(2) a composite spin-orbital excitation resulting from SOC.
Hence, our results give experimental support for \Ca\ being in a spin-orbit coupled band-Mott insulating phase. Thus, it provides an experimental unification of the band-Mott~\cite{SutterNatComm2017a,GorelovPRL2010} and van Vleck-type Mott~\cite{KhaliullinPRL2013,KhaliullinPRB2014} insulator scenarios.

\textit{Methods.} 
High quality single crystals of \Ca\ were grown by the floating zone techniques~\cite{FukazawaPhysB00,snakatsujiJSSCHEM2001}. X-ray absorption spectroscopy (XAS) and RIXS \cite{amentRMP2011}
were carried out at the ADRESS beamline~\cite{strocov2010high,ghiringhelliREVSCIINS2006}  at the Swiss Light Source (SLS). 
The scattering geometry is indicated in Fig.~\ref{fig:fig1n}(a). A fixed angle of 130$^\circ$ between incident light and scattered light was used. In-plane momentum is varied by controlling the incident photon angle $\theta$ 
shown in Fig.~\ref{fig:fig1n}(a). Grazing and normal incidence conditions refers to $\theta\approx 90^\circ$ and $0^\circ$ respectively.
Linear vertical(LV) and horizontal (LH) light polarizations were used to probe the oxygen $K$-edge at which an energy resolution
of 29~meV or better (half width at half maximum) was obtained. 
Despite the orthorhombic low-temperature (S-Pbca) crystal structure of \Ca,  we indicate momenta $Q=(h,k,\ell)$ 
using tetragonal 
notation in reciprocal lattice units, 
with $a\approx b=3.84$~\AA\ and $c\approx 11.95$~\AA. Furthermore, since \Ca\ is a quasi two-dimensional system, we consider only the 
planar component $Q_{||}=(h,k)$ involved in the RIXS process. Elastic scattering is throughout this work modeled by a Voigt-lineshape, allowing subtraction of this component. The presented data is collected at $T=16$~K unless otherwise indicated.\\[2mm]

\textit{Results.} 
 XAS spectra recorded with LH light polarization near normal  and grazing  incidence conditions are shown in Fig.~\ref{fig:fig1n}(b). Good agreement with previous published XAS experiments~\cite{malvestutoPRB2011,mizokawaPRL2001,malvestuto2013nature}
is found when overlap in temperature, light polarization and incident angle allows for a comparison. 
Common to single layer perovskite structured transition metal oxide materials~\cite{ctchenPRL1991,salaPRB2014,mizokawaPRL2001,malvestutoPRB2011},  
the planar oxygen absorption resonance is found 1--2~eV above that of the apical site. 
As previously reported~\cite{malvestutoPRB2011,mizokawaPRL2001,FatuzzoPRB2015},  the apical and planar 
oxygen $K$-edge peaks are found at $\sim$529.1~eV and $\sim$529.8~eV [see Fig.~\ref{fig:fig1n}(b)]. 
These resonances stem from hybridization of the oxygen $p$-bands with the ruthenium $t_{2g}$ states whereas
the resonances at higher photon energies are related to hybridization with unoccupied $e_{g}$ states. 

\begin{figure}
\begin{center}
\includegraphics[width=0.45\textwidth]{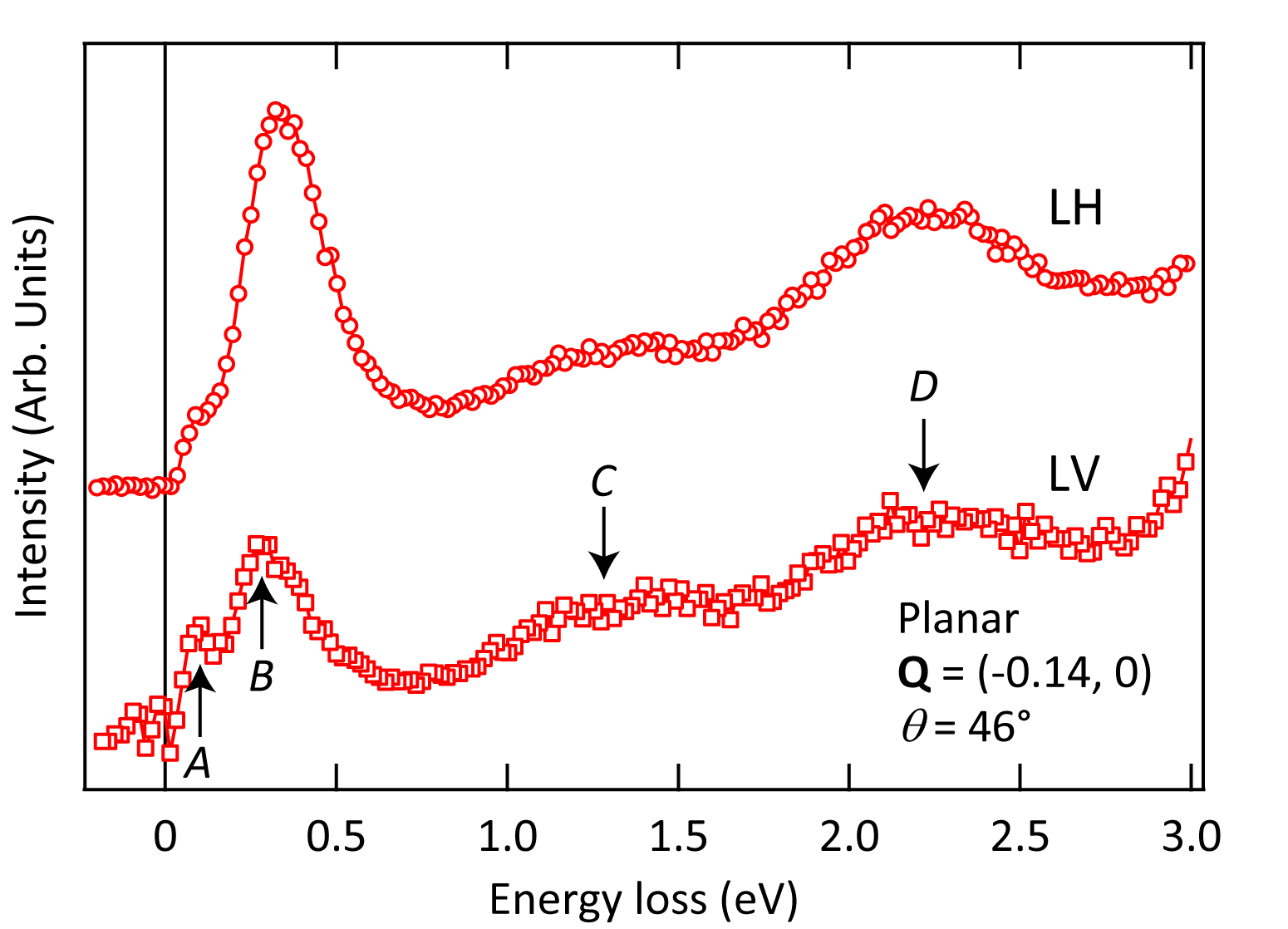}
\end{center}
\caption{(color online) Planar RIXS spectra, with elastic scattering subtracted and recorded with LH and LV light polarization for incident angle (momentum transfer) as indicated.  Vertical arrows indicate the four excitations labelled \A, \B, \C, and \D. For clarity, the spectra are given an arbitrary vertical shift }

\label{fig:fig2n}
 \end{figure}

\begin{figure*}
\begin{center}
\includegraphics[width=0.99\textwidth]{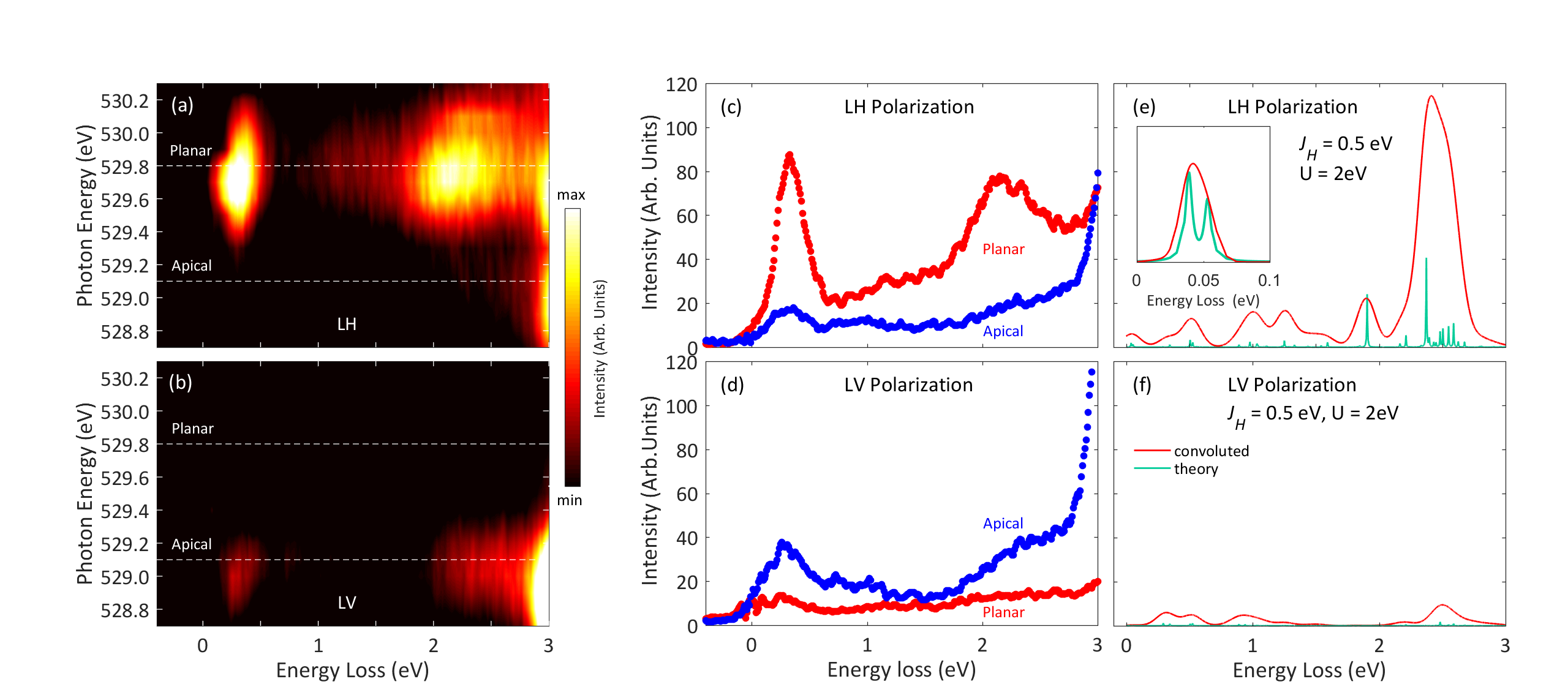}
\end{center}
\caption{(color online) Polarization dependence of the RIXS spectra versus incident photon-energy. (a) and (b) RIXS response, in false intensity scale, as a function of energy loss and incident photon energy of LH  and LV light polarization for grazing incidence condition as indicated. Horizontal dashed lines show the positions of the apical and planar resonances obtained from XAS.
(c) and (d) RIXS spectra, with the elastic response subtracted, at the apical (blue) and planar (red) oxygen resonances for the respective light polarizations. (e) and (f) calculated RIXS spectra for the planar site with respect to linear horizontal (c) and vertical (d) light polarization, see text for a detailed explanation of the model. Green lines indicate the expected excitations and solid red line is obtained by Gaussian convolution to mimic instrumental resolution. A standard deviation $\sigma=70$~meV was applied 
in (e) and (f) whereas $\sigma=7$~meV was used for the inset that 
displays a zoom on the lowest excitations at around $40$~meV.
 }

\label{fig:fig3n}
 \end{figure*}

\begin{figure*}
\begin{center}
\includegraphics[width=0.99\textwidth]{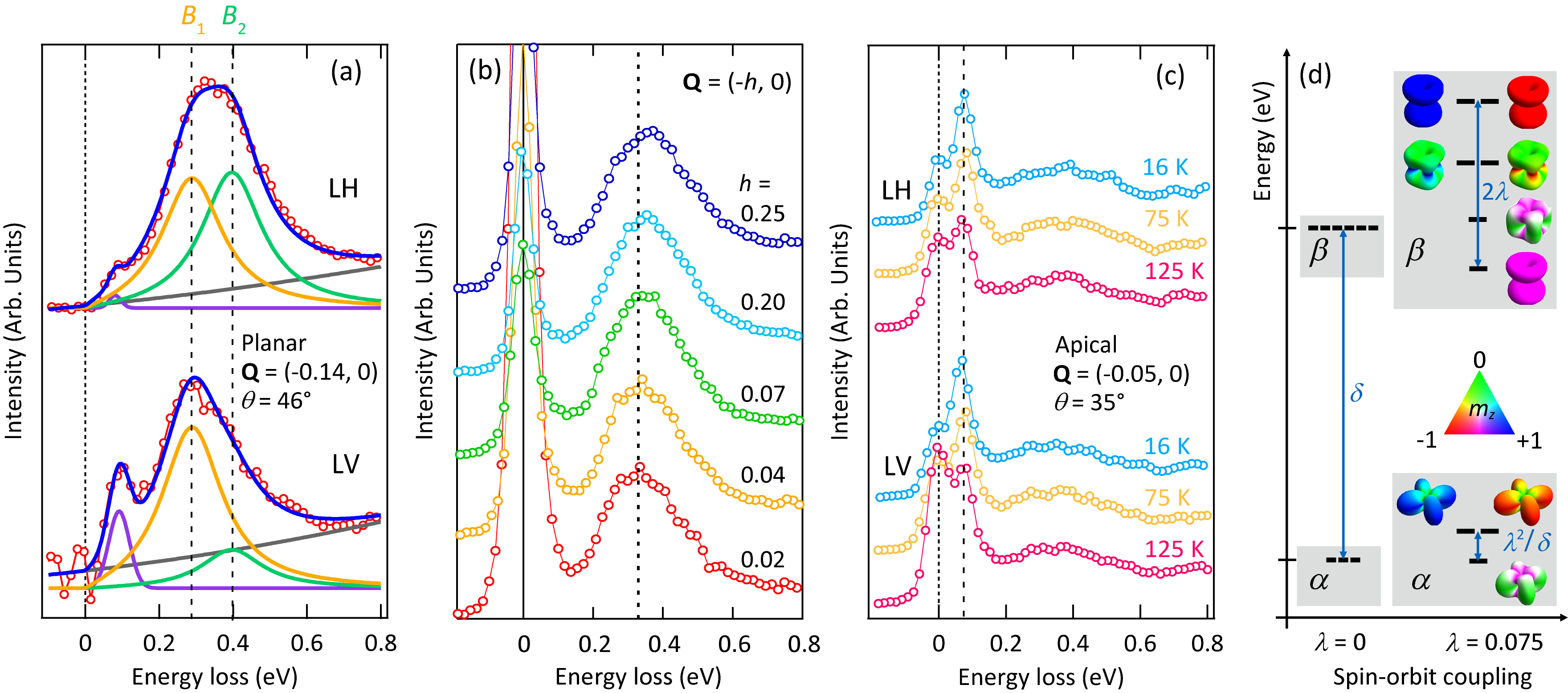}
\end{center}
\caption{(color online)  
(a) Same spectra as in Fig.~2 displayed with emphasis on the low-energy excitations. Solid blue lines are a four component fit including a smoothly growing background (gray lines -- second-order polynomial form), the 80 meV excitation (purple line -- Gaussian lineshape) and two modeswith damped harmonic oscillator lineshape~\cite{le2011intense,monney2016resonant,LamsalPRB2016} labelled $B_1$ (yellow) and $B_2$ (green) for the excitation at 400 meV.
The position and width of $B_1$ and $B_2$ where assumed identical for LV and LH polarization. Peak amplitudes by contrast were left as open fit parameters irrespectively of light polarization. (b) RIXS spectra at planar resonance recorded using linear horizontal light polarization  as a function 
 in-plane momenta as
indicated.  (c) Apical RIXS spectra recorded with LV and LH polarization near normal incidence for temperatures as indicated. Lines in (b) and (c) are guides to the eye.  
(d) schematics of low-lying energy levels of an interacting model for a single ruthenium site for spin-orbit coupling $\lambda$ set to zero (left) and to the physical value in \Ca\ (right). 
With four electrons, one of the orbitals \dxy, \dxz, \dyz\ is doubly occupied and the two singly occupied electrons are in a spin-triplet state.  
Finite spin-orbit coupling 
lifts the degeneracies of the two sectors denoted $\alpha$ and $\beta$. The character of the doubly occupied orbital is displayed along with a color scale indicating the directional dependence of the total spin $m_z$ moment. }
\label{fig:fig4n}
 \end{figure*}

In Fig.~\ref{fig:fig2n}, four RIXS distinct excitations -- labelled \A, \B, \C, and \D\ -- with approximate 
energy losses of 0.08, 0.4, 1.3 and 2.4 eV are displayed in addition to elastic scattering and $dd$-excitaions in the 3-5 eV range [see Figs.~\ref{fig:fig1n}(c--d)]. Only the \B\ excitation (at $\sim0.4$~eV) has previously been discussed in Ref.~\onlinecite{FatuzzoPRB2015}. 
The amplitude 
of these excitations strongly depend on incident light angle and polarization. These matrix elements are furthermore different on the apical and planar 
resonances. All four excitations are therefore not necessarily visible in a single spectrum -- as in Fig.~\ref{fig:fig2n}. We start by discussing the two most intense excitations \B\ and \D. Plotting the photon-energy dependent RIXS response (Fig.~\ref{fig:fig3n}) for grazing incident condition, these two excitations are the most prominent features in the spectra. They are particularly intense on the planar oxygen $K$-edge resonance for LH-polarization. 
Interestingly, these excitations are 
virtually ``turned off" when the light polarization is switched to linear vertical (LV) polarization. The opposite polarization dependence is observed on the apical 
site where the excitations are observed for LV and suppressed for LH polarized light.  
The same light polarization analysis for an incident angle between grazing and normal incidence is shown in Fig.~\ref{fig:fig2n} and Fig.~\ref{fig:fig4n}(a) for the planar resonance. It reveals several important insights. (1) The lineshape of the \B\ excitation is strongly dependent on the incident light polarization. In fact, the peak maximum depends on light polarization [Fig.~\ref{fig:fig4n}(a)]. (2) The \D\ excitation is stronger for the grazing incidence and generally weaker in the LV channel. By contrast, the \C\ excitation  is more visible with LV polarization [Fig. 2]. (3) The same is true for the \A\ excitation: on the planar resonance, it is barely resolvable  with LH light, but appears clearly in the LV channel. In Fig.~\ref{fig:fig4n}(c), it is demonstarted how the \A\ excitation appears  in both the LH and LV channels on the apical resonance (near normal incidence). (4) The linewidth of the \A\ excitation is essentially resolution limited and hence much sharper than that of \B. The implications of this observation will be discussed in greater detail below.

\begin{figure*}
\begin{center}
\includegraphics[width=0.99\textwidth]{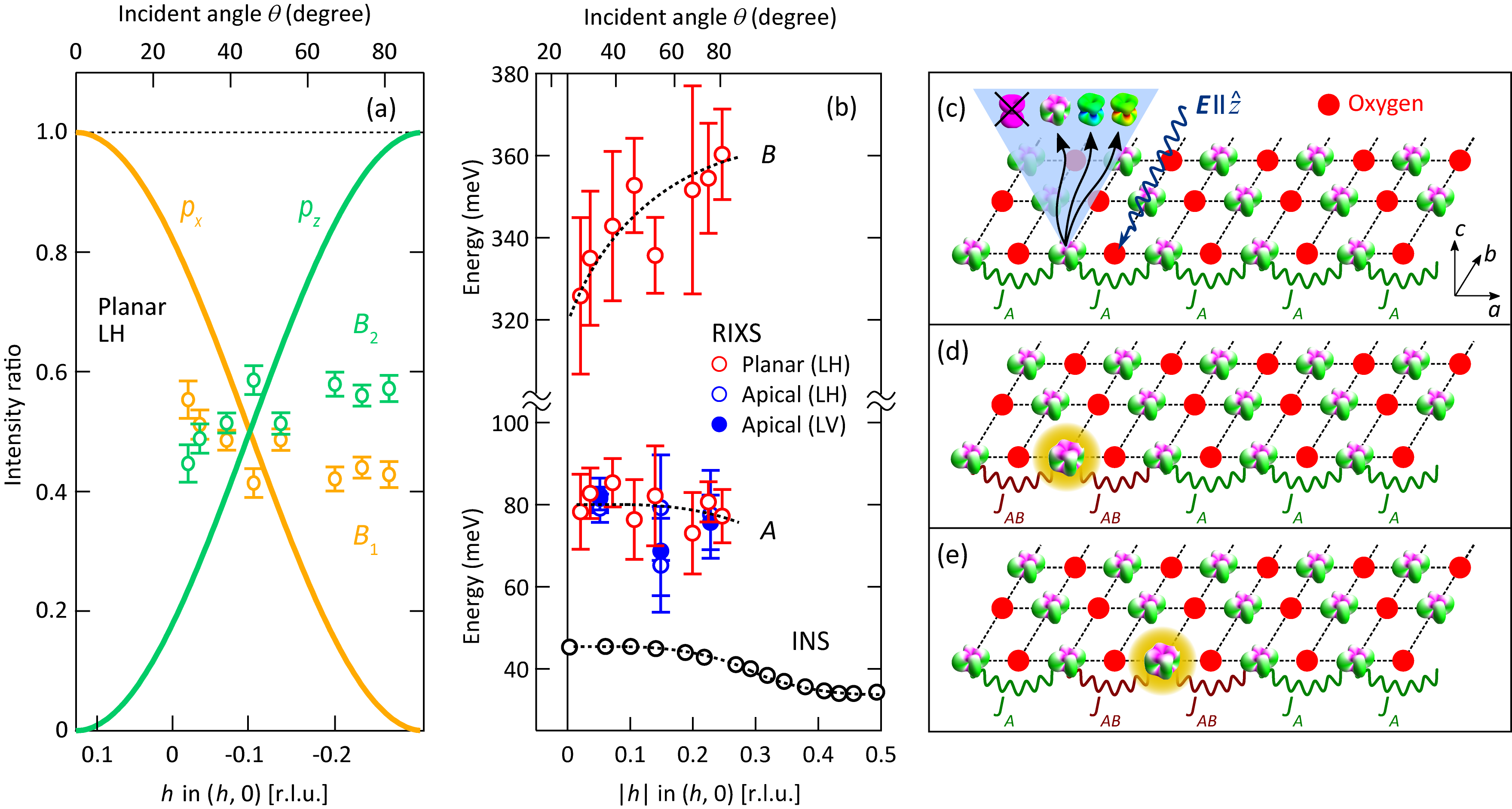}
\end{center}
\caption{(color online) 
Momentum dependent analysis of the low-energy excitations. (a) Peak amplitudes of the assumed $\B_1$ (yellow) and $\B_2$ (green) modes extracted from fits (with fixed peak widths and positions -- see text) as shown in Fig.~\ref{fig:fig4n}(a).
Corresponding solid cosine squared lines indicate, for LH, the expected coupling to the $p_x$ and $p_z$ oxygen orbitals. 
(b) Extracted dispersion along the $Q=(h,0)$-direction of the $A$ and $B$  excitations in meV as shown before in panel (a) of Fig.~\ref{fig:fig4n}. The dispersion is defined by the peak maxima which in case of the $B$-excitation is derived from a derivative of the spectra. Error bars indicate standard deviations $3\sigma$ and $\sigma$ for the $A$ and $B$, respectively.  
Comparison to the spin excitation branch observed by neutron scattering  (reproduced from Ref.~\cite{jain2017higgs}) along the same direction is also shown in panel (b). Dashed lines are guides to the eye. 
 (c) Schematic of an oxygen $K$-edge RIXS process creating a local excitation between the $\alpha$ and $\beta$ sectors (see Fig.~\ref{fig:fig4n}).  (d) and (e) illustrate  propagation of the  spin-orbital excitation. The nearest neighbor couplings are denoted $J_A$ between sites in the $\alpha$ sector and $J_{AB}$ between  $\alpha$ and $\beta$ sectors.}
\label{fig:fig5n}
 \end{figure*}

We now turn to discuss temperature and momentum dependence of the \A\ and \B\ excitations. As evident from Fig.~\ref{fig:fig4n}(c), 
both the \A\ and \B\ excitations persist into the paramagnetic phase. The momentum dependence -- along the $(h,0)$ (Ru-O bond) direction -- of planar spectra recorded with LH polarization is shown in Fig.~\ref{fig:fig4n}(b). 
The peak maximum position, extracted from fitting the derivative of the spectra, reveals a weak 
momentum dependence consistent with a dispersive \B\ sector.  The extracted momentum dispersion of the excitation is reported  [Fig.~\ref{fig:fig5n}~(b)] with a minimum at the zone center.  In comparison no dispersion of the \A\ excitation could be resolved within the applied energy resolution.   
For completeness, the RIXS data are compared with the 
 amplitude spin excitation mode reported by inelastic neutron scattering (INS)\cite{jain2017higgs}.
\\[2mm]


\textit{Discussion.}
The exact nature of the Mott insulating state of \Ca\ has long been debated. 
Different theoretical models have been put forward \cite{Anisimov2002,LiebschPRL2007,GorelovPRL2010,CuocoPRB06a,CuocoPRB06b}. Some of them suggest that all $t_{2g}$ orbitals are involved in the Mott transition. 
Other models proposed that crystal fields drive 
the \dxy\ states band insulating and the Mott physics is induced on the resulting half-filled $d_{xz/yz}$ 
bands~\cite{LiebschPRL2007}. A recent ARPES study of the paramagnetic Mott insulating state supports this combined band-Mott
insulating scenario~\cite{SutterNatComm2017a}. This conclusion was reached by visual comparison of the measured and calculated spectral functions for different scenarios. 

Based on this development, it is interesting to evaluate the implications of the band-Mott scenario on the XAS and RIXS spectra. 
When the \dxy\ orbital is (almost) completely occupied, it is inaccessible to the XAS processes that require unoccupied states. 
Therefore, \dxz\ and \dyz\ are the main active $t_{2g}$ states available for absorption.
The XAS spectra, shown in Fig.~\ref{fig:fig1n}(b), are in perfect accordance with this picture.
For example, near grazing incident condition using LH polarization, the core electron is promoted into the $p_z$ oxygen orbital that at the planar-site hybridizes with $d_{xz/yz}$. Indeed, a pronounced response is observed at the planar oxygen $K$-edge resonance, whereas the intensity at the apical resonance is strongly suppressed. Further, changing to normal incidence (keeping LH polarization), 
the core electrons are promoted into the oxygen $p_x$ orbital that at the planar site 
hybridizes with \dxy\ and at the apical site with \dxz . As shown in Fig.~\ref{fig:fig1n}, the XAS response flips to the apical resonance. Our XAS results thus suggest that the unoccupied $t_{2g}$ states have predominant $d_{xz/yz}$ character.

The intensities of the RIXS spectra naturally follow the polarization dependence of the XAS response.
We notice that the two excitations \B\ and \D\ observed at 0.4~eV and 2.2~eV respectively are most pronounced near 
 grazing incidence with LH polarization at the planar oxygen resonance. This is exactly 
where the oxygen hybridization with $d_{xz/yz}$ is optimized. 
We thus conclude that the excitations are intimately linked to the unoccupied $d_{xz/yz}$ states. The \A\ excitation by contrast is observed on both the apical and planar resonances with both LH and LV polarization. This suggests that the orbital character of this excitation involves a mixture of \dxy\ and $d_{xz/yz}$ states.  

To connect our data with a microscopic physical picture, we  computed the RIXS response for an interacting model of \Ca.
Within the fast collision approximation~\cite{AmentPRB2007,amentRMP2011} the RIXS cross-section for exciting an electron from the oxygen $1s$ level into a $2p_k$ level,  with $k=x,y,z$, is given by
\begin{equation}
I^{\mathrm{p}/ \mathrm{a}}_{p_k} \propto \sum_{m}\sum_{i}\left|\left\langle m\left|\hat{n}^{\mathrm{p}/\mathrm{a}}_{p_k,i}\right|0\right\rangle\right|^2\delta[\omega-(E_m-E_{\mathrm{0}})],
\label{eq: RIXS cross section}
\end{equation}
where the operator $\hat{n}^{\mathrm{p}/\mathrm{a}}_{p_k,i}$ measures the \emph{hole} density of oxygen $p_k$ orbitals on all planar (p) or apical (a) oxygen sites surrounding the ruthenium site $i$, $|0\rangle$ is the ground state with energy $E_{\mathrm{0}}$ and the sum $m$ 
runs over all excited states $|m\rangle$ with energies $E_m$.

To discuss the spectra presented in Figs.~\ref{fig:fig1n} and~\ref{fig:fig3n}, we model a cluster of two ruthenium sites connected by one planar oxygen site (see supplementary information).
The ruthenium site Hamiltonian consists of three terms: 
(1) Crystal field splitting $\delta$ between the $d_{xy}$  and  $d_{xz}$, $d_{yz}$ orbitals, 
(2)  SOC $\lambda$, and
(3)  Coulomb interaction, which is expanded into intra-orbital and inter-orbital Hubbard interactions of strengths $U$ and $(U-5J_{\mathrm{H}}/2)$, respectively. Inter-orbital Hund's coupling as well as the pair-hopping term, are both of strength $J_{\mathrm{H}}$. 
To evaluate the model,  material specific
 values $\delta = 0.3$~eV, $\lambda=0.075$~eV, $U=2$~eV, and $J_{\mathrm{H}}=0.5$~eV~\cite{mizokawaPRL2001,veenstraPRL2014,FatuzzoPRB2015}, are used.
 Similar values of $\delta$, $U$ and $J_{\mathrm{H}}$ have been used for DMFT calculations~\cite{SutterNatComm2017a} of \Ca\ and 
 the ratio $\delta/(2\lambda)=2$ is comparable to what was used in modelling the spin-excitation dispersion observed by neutron scattering~\cite{jain2017higgs}.
 We stress that qualitatively, the model is not very sensitive to the exact set of parameters. 
 Although the ratio of spectral weight between the low- and high-energy excitations for horizontal polarization is different from the data, 
our results presented in Figs.~\ref{fig:fig3n}(e--f), reproduce qualitatively the experimental spectra, in particular the excitation at about $2.2$~eV as well as the polarization dependence of the spectral weight. 
We point out that the spectral features at about 1~eV and 2~eV arise from single and double singlet-triplet excitations at the ruthenium site with an energy of $2 J_{\mathrm{H}}$ and $4 J_{\mathrm{H}}$, respectively. This thus provides an explanation for the observed $C$ and $D$ excitations. Such modes are spin-orbit activated when mixing $d^4$ with $d^3$ or $d^5$ states 
and represent the lowest energy singlet-triplet excitations 
when the total number of doubly occupied orbitals at the ruthenium sites is held fixed  (see Supplemental Information).

To elucidate the nature of the low-energy excitations, we concentrate on the local electronic structure at a single ruthenium site.
The low-energy configurations have four electrons ($d^4$), one doubly occupied orbital (doublon), and the two other electrons in a spin-triplet state.
For $\lambda=0$, the model has a three-fold degenerate ground state manifold $\alpha$ with a doublon in the \dxy\ orbital. The lowest-lying excitation sector $\beta$ is six-fold degenerate at energy $E_{\alpha\beta}=\delta$, with the doublon in the \dxz\ and \dyz\ orbitals, see Fig.~\ref{fig:fig4n}(d). Finite spin-orbit coupling has two effects: (i) it lifts the degeneracies of the  $\alpha$ and $\beta$ states by introducing a splitting of about $\lambda^2/\delta$ and $2\lambda$, respectively, and crucially (ii) it mixes the orbital character of the doublon state. The $\beta$ states thus correspond to a spin-orbit{\textcolor{red}{al}} excitation. The splitting of  $\alpha$ states gives rise to low-energy excitations that have been studied using neutron scattering~\cite{jain2017higgs} [reproduced in Fig.~\ref{fig:fig4n}(b)]. Just as the $\beta$ sector has an expected internal orbital structure, it was recently demonstrated by Raman spectroscopy that the low-energy $\alpha$ sector also consists of multiple excitations. In fact, a Raman study also revealed two excitations around 80 meV and associated them with two-Higgs and two-magnon scattering modes~\cite{Souliou2017}. 
Although optical 80-100 meV phonon modes are not uncommon in transition metal oxides, the Raman study~\cite{Souliou2017} suggests that our $A$-excitation is of magnetic origin. 
To this end, we stress that our model spectra
shown in Fig.~\ref{fig:fig3n}(e) display low-energy modes with maximum intensity at about 40 meV. It can be assigned to the amplitude and phase excitations arising from the effective $J_{\mathrm{eff}}=1$ configurations
    in the $\alpha$ sector.   Our cluster analysis, by construction, does not allow to
    obtain multiple amplitude excitations associated to the interacting
   $J_{\mathrm{eff}}=1$. A two-scattering mode (i.e. nearby 80~meV) is, however,  by principle expected and would emerge in a larger cluster calculation and eventually considering the RIXS cross-section at the oxygen K-edge beyond the Fast Collision approximation.
 The predicted 40 meV magnetic mode, observed by neutron scattering, should in principle also enter into the RIXS cross-section. It is, however, not observed in our experiment due to the finite energy resolution. 


The broadness of the \B\ excitation and its light polarization dependence may be interpreted as a consequence of the internal structure of the $\beta$ sector. In fact, it is possible to fit the \B\ excitation 
with two internal levels 
labeled $B_1$ and $B_2$. Keeping identical linewidths and fixed peak positions, the fits describe both the light polarization dependence [Fig.~\ref{fig:fig4n}(a)] and the momentum dependence by fitting the peak amplitudes. In particular, the observed light polarization dependence for a fixed incident angle (momentum transfer) strongly suggests that the \B\ excitation indeed has an internal orbital structure.
This structure constitutes a clear difference to the single exciton excitation found on the strongly spin-orbit coupled Mott insulator Sr$_2$IrO$_4$~\cite{KimNatComm14}.

The internal structure raises the need to clarify whether the observed dispersion of the \B\ excitation is a result of matrix element variation of this internal structure.
By varying the incident angle (momentum transfer) with LH polarization, a switch between $p_z$ (grazing incidence) and $p_x$ (normal incidence) is effectuated. The expected sensitivity to the  $p_z$ and $p_x$ oxygen orbitals is shown in Fig.~\ref{fig:fig5n}(a). In the same figure, the fitted peak amplitudes of the $B_1$ and $B_2$ levels are shown. As they vary only weakly with momentum, the matrix element effect does not provide a plausible explanation for the observed dispersion. We thus conclude that the dispersion is intrinsic which in turn indicates an itinerant nature of this sector.

When charge hopping between ruthenium sites is reinstated, these local spin-orbit excitons can propagate. Indeed, they acquire a dispersion through virtual processes involving $d^3$-$d^5$ excitations on neighboring ruthenium sites. Within second order perturbation theory, one obtains an estimate of the bandwidth of the spin orbital excitation in the range of $\sim30$--$40$~meV, using $t\sim0.25$--$0.3$~eV \cite{GorelovPRL2010} for the ruthenium inter-site hopping and $\delta = 0.28$~eV, $\lambda=0.075$~eV, $U=2.2$~eV, and $J_{\mathrm{H}}=0.4$~eV for the other electronic parameters (see Supplemental Information). It is furthermore expected that excitations in the $\alpha$-sector exhibit a weaker dispersion due to the smaller exchange amplitude between the $J_\mathrm{eff}=1$ modes. 
Our model thus account qualitatively for the fact that the \B\ excitation disperse about 30 meV over half a zone whereas the $A$ excitation, according to inelastic neutron scattering~\cite{jain2017higgs} [see Fig.~\ref{fig:fig5n}(b)], disperse no more than 20 meV over the entire zone.



\textit{Conclusions and Outlook:}
In summary, we have carried out a comprehensive oxygen $K$-edge RIXS study of \Ca. 
We demonstrate that the strong light polarization dependence of the signal  is a direct manifestation of the band-Mott insulating nature of \Ca. The hybridization between oxygen $p$ and ruthenium $d$ states  thus primarily involves the $d_{xz/yz}$ orbitals. 
Although the system has a modest SOC, it is a crucial element 
to explain our observations. Most importantly it allows for a distinct set of propagating low-energy ($0.4$~eV) excitations with a spin-orbital character. The spin-orbit coupling is also relevant for activating the high-energy 
($\sim$2.2~eV) non-dispersive excitations by the RIXS process, achieved by local conversion of triplet into singlet states. For realistic values of crystal fields, Hund's coupling $J_{\mathrm{H}}$, Coulomb interaction $U$, and SOC, all salient features 
of the RIXS spectra were captured with minimal theoretical modeling.
Our results demonstrate that \Ca\ is a Mott insulator with a paradigmatic 
competition between SOC and crystal field energy scales.
Combining RIXS data and theoretical modelling, we unveiled how spin-orbital entangled excitations manifest within a spin-orbit coupled band-Mott insulator.  
For future studies it would be of great interest to further resolve the internal structure 
of the low-energy excitations. 
We envision two
different pillars of experimental strategies that alone or in combination would
allow further insight.  (1) As synchrotrons are upgrading for diffraction limited experiments, 
flux and resolution at the oxygen $K$-edge will improve. 
In particular enhanced energy resolution, in combination 
with the light polarization analysis put forward here, would 
allow us to study important information on 
different orbital characters of these excitations. With gains in energy resolution, 
the RIXS technique is also going to enter further into the spin-excitation sector. (2) Direct high-resolution RIXS experiments on the ruthenium $L$-edge 
is another promising avenue.     \\[2mm]

\textbf{Acknowlegdements}
This work was performed at the ADRESS beamline of the SLS at the Paul Scherrer Institut, Villigen PSI, Switzerland. We thank the ADRESS beamline staff for technical support.
 L.D., J.C., M.H., O.I., M.D, and F.S. acknowledge support by the Swiss National Science Foundation through 
 the  SINERGIA network Mott Physics beyond the Heisenberg Model and  under grant numbers BSSGI0-155873, 200021-169061, P2FRP2-171824, 
 and 200021L-141325.
 P.O.V is also funded by SINERGIA.
 L.D. is partially funded by a Swiss Government  PhD  excellence scholarship. J.C. acknowledges the CNR Short Term Mobility Program (STM 2016) for partial financial support.
J.P and T.S acknowledge financial support through the Dysenos AG by Kabelwerke Brugg AG Holding, Fachhochschule Nordwestschweiz and the Paul Scherrer Institute. 
J.P acknowledges financial support by the Swiss Mational Science Foundation Early Postdoc. Mobility Fellowship project number P2FRP-171824.
P.O.V. acknowledges financial support from the European Community Seventh Framework Programme (FP7/2007-2013) under Grant Agreement No. 290605 (PSIFELLOW/COFUND).
\\


%

\end{document}